\documentclass[letter,twocolumn,showpacs,amsmath,amssymb,aps,prl,superscriptaddress]{revtex4}
\usepackage{graphicx}
\input{babarsym.tex}

\begin{document}

{\pagestyle{empty}

\par\vskip 5cm
\begin{widetext}

\title{Observation of the decay $\Bzb\to\Lambda_c^+\antiproton\piz$}
%
\author{B.~Aubert}
\author{Y.~Karyotakis}
\author{J.~P.~Lees}
\author{V.~Poireau}
\author{E.~Prencipe}
\author{X.~Prudent}
\author{V.~Tisserand}
\affiliation{Laboratoire d'Annecy-le-Vieux de Physique des Particules (LAPP), Universit\'e de Savoie, CNRS/IN2P3,  F-74941 Annecy-Le-Vieux, France}
\author{J.~Garra~Tico}
\author{E.~Grauges}
\affiliation{Universitat de Barcelona, Facultat de Fisica, Departament ECM, E-08028 Barcelona, Spain }
\author{M.~Martinelli$^{ab}$}
\author{A.~Palano$^{ab}$ }
\author{M.~Pappagallo$^{ab}$ }
\affiliation{INFN Sezione di Bari$^{a}$; Dipartimento di Fisica, Universit\`a di Bari$^{b}$, I-70126 Bari, Italy }
% ...
\collaboration{The \babar\ Collaboration}
\noaffiliation

{\babar}-PUB-09/28 \\
SLAC-PUB-14189\\

\begin{abstract} 
 In a sample of 467 million \BB pairs collected with the \babar~detector at the \pep2 collider at SLAC we have observed the decay 
$\Bzb\rightarrow\Lambda_c^+\antiproton\piz$ and measured the branching fraction to be 
$(1.94\pm0.17 \pm 0.14 \pm 0.50)\times 10^{-4}$,
 where the uncertainties are statistical, systematic, and the uncertainty on the $\Lambda_c^+\rightarrow\proton\Km\pip$ branching fraction, respectively. 
We determine an upper limit of $1.5\times 10^{-6}$ at $90\%$ C.L. for the product branching fraction $\BR(\Bzb\to\Sigma_c^+(2455)\antiproton)\times\BR(\Lambda_c^+\to\proton\Km\pip)$. 
Furthermore, we observe an enhancement at the threshold of the invariant mass of the baryon-antibaryon pair.
\end{abstract} 
\date{\today}
\maketitle
\end{widetext}
}

{~}
\clearpage

Although approximately $7\%$ of $B$-meson decays have baryons in the final state, presently the sum of all measured branching fractions of exclusive baryonic $B$ decays is only about $1\%$  \cite{PDG}. 
$\B$ mesons decay dominantly via $\b\to\c$ transitions, hence decays to baryons should be 
dominated by charm baryon production or a charmed meson accompanied by non-charmed baryons. 
Both types of decays have been observed \cite{ref:dmeson1,ref:dmeson2}, and are found to have comparable branching fractions for decays to final states with the same multiplicity.

In baryonic \B decays and in baryon production in general, enhancements at the threshold for the baryon-antibaryon invariant mass have been observed \cite{ref:dmeson2,ref:enh2}. 
This may indicate resonances near threshold or another mechanism for enhanced production of baryon-antibaryon pairs.
This threshold enhancement may also explain the increase in branching fraction with final state multiplicity and the apparent suppression of two-body decays to baryons \cite{PDG,ref:theorie}.

The mechanisms of baryon production in heavy meson decays are poorly understood, and studies of exclusive decays may provide insight into different decay mechanisms.
As will be discussed below, isospin relations will also help distinguish different primary  processes.

In this paper, we present a study of  the decay $\Bzb\to\Lambda_c^+\antiproton\piz$ \footnote{Throughout this paper charge conjugate modes are always implied.} and measure its branching fraction.
The CLEO collaboration  previously set an upper limit of $\BR(\Bzb\rightarrow\Lambda_c^+\antiproton\piz)<5.9\times 10^{-4}$ based on an 
integrated luminosity of 2.39 \invfb \cite{cleo1996}. For the isospin-related decay, $\Bm\to\Lambda_c^+\antiproton\pim$, several measurements of the branching fraction have been performed \cite{ref:isospin1,ref:isospin2}.
The recent \babar~measurement gives $(3.38\pm0.12\pm0.12\pm0.88)\times 10^{-4}$ \cite{ref:stephanie}, a value that is significantly higher than earlier measurements ($4.3\sigma$ deviation). 
The last and dominant error is due to the uncertainty in the $\Lambda_c^+\to\proton\Km\pip$ branching fraction, common to all measurements.

While the $\Bm\to\Lambda_c^+\antiproton\pim$ final state can only have an isospin $I$ of $3/2$, $\Bzb\to\Lambda_c^+\antiproton\piz$  can also have $I=1/2$.
If both decays proceed via the same weak decay mechanism, $I=3/2$, the ratio of the partial decay widths of \Bzb to \Bm should be $2/3$.
However, it is also possible that the decay mechanisms are different. Thus a deviation of the ratio of partial decay widths from $2/3$ would suggest a contribution from the $I=1/2$ final state to the $\Bzb\to\Lambda_c^+\antiproton\piz$ decay or a contribution from the decay process where the $\pim$ is coming from the $W$ in the $\Bm\to\Lambda_c^+\antiproton\pim$ decay.

This analysis is based on a dataset of about 426\invfb corresponding to  467 million \BB pairs. These data were collected with the \babar\ detector
at the \pep2\ asymmetric-energy \epem\ collider with a center-of-mass energy, $\sqrt{s}$, at the \FourS resonance mass. An additional sample of 44.5\invfb, collected 40\mev below the 
mass of the \FourS resonance, are used to study the continuum background $\epem\to\qqbar$, where $\q=\u,\d,\s$ or $\c$. 

The signal efficiency is determined using a detailed \textsc{Geant4} \cite{geant4} Monte Carlo (MC) simulation of the \babar\ detector that generates MC events uniformly in the $\Lambda_c^+\antiproton\piz$ phase space. MC events are also used to study the background contributions.

The \babar\ detector is described in detail elsewhere~\cite{ref:babar}. 
Charged particles are distinguished and their momenta measured in the tracking system consisting of a five-layer double-sided silicon vertex tracker (SVT) and a 
40-layer drift chamber (DCH). An internally reflecting ring imaging Cherenkov detector (DIRC) is also used to distinguish charged particles and a CsI(Tl) electromagnetic calorimeter (EMC) is used to detect photons.

Likelihood ratios based on information from SVT, DCH and DIRC are used to identify protons and kaons.
The efficiency for the kaon selection is around $90\%$ while the rate for misidentifying pions and protons as  kaons varies between $5\%$ and $10\%$, depending on track momentum.
The identification efficiency for the proton selection is greater than $90\%$ while the misidentification rate of identifying kaons and pions as protons 
varies between $3\%$ and $15\%$, depending on track momentum. 

Two photons are selected as electromagnetic showers in the EMC with the expected shape and are combined to form a \piz candidate where the photon with the lower energy must have an energy greater than $60\mev$, while the second photon must have an energy greater than $100\mev$.
The invariant mass of the $\gamma\gamma$ combination is required to be between 120\mevcc and 145\mevcc.
The $\Lambda_c^+$ candidates are reconstructed in the decay mode $\Lambda_c^+\to\proton\Km\pip$, and a fit with geometric constraint applied to the common vertex must have a $\chi^2$ probability greater than $0.1\%$. 
The invariant $\proton\Km\pip$ mass must be within $2.5\sigma$ of the fitted peak of the mass distribution, $2.276<m(\proton\Km\pip)<2.296 \gevcc$.
The $\Lambda_c^+$ and \piz candidates are then combined with a $\antiproton$ candidate in a fit using kinematic constraints to form a \Bzb candidate.
In the fit the mass of the $\proton\Km\pip$ candidate is constrained to the mass of the $\Lambda_c^+$ and the mass of the $\gamma\gamma$ combination to the mass of the \piz \cite{PDG}. 
The $\chi^2$ probability of this fit must be greater than $0.1\%$.

\begin{figure}
\includegraphics[width=\linewidth]{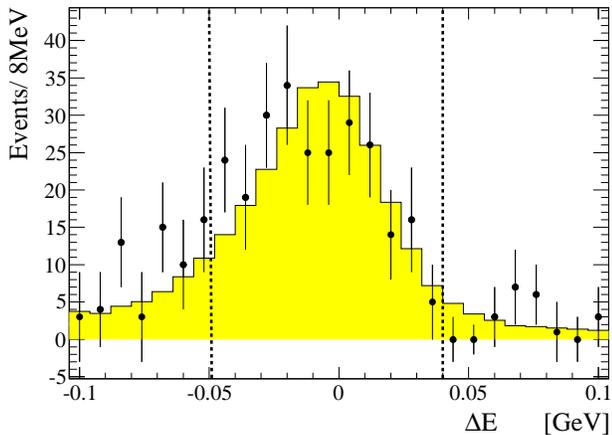}
\caption{$\DeltaE$ distribution for data signal events after all selection cuts (data points)
and signal MC events (histogram) normalized to the number of data signal events; signal events are obtained from binwise \mes fits; dashed lines show the range used for \mes distributions. }
\label{fig:deltaE}
\end{figure}

The analysis makes use of two almost independent kinematic variables, $\Delta E$ and $\mes$, where
$\DeltaE=E^*_B-\sqrt{s}/2$ is the difference of the reconstructed energy $E^*_B$
and half of $\sqrt{s}$ in the \epem center of mass frame (CMS). The other variable is $\mes=\sqrt{(s/2+{\bf p_0\cdot p_B})^2/E_0^2-{\bf p_B^2}}$ 
where $(E_0,{\bf p_0})$ is the four momentum of the \epem system and ${\bf p_B}$ is the \B candidate momentum, both measured in the laboratory frame. 
The \mes distribution for signal events peaks at the $B$ mass and the distribution of \DeltaE for signal events is centered around zero. 
Candidates arising from other $B$ decays, with more final-state particles, such as $\Bzb\to\Lambda_c^+\antiproton\pip\pim$, are shifted 
to negative values of \DeltaE.
Conversely, candidates arising from $B$ decays with fewer final-state particles, such as $\Bzb\to\Lambda_c^+\antiproton$, are shifted to positive values.
To suppress these decays only candidates with $-50\mev<\DeltaE<40\mev$ are selected.

A considerable background comes from $\Bm\to\Lambda_c^+\antiproton\pim$ decays, and in particular from the $\Bm\to\Sigma_c^0(2455)\antiproton$, $\Sigma_c^0(2455)\to\Lambda_c^+\pim$ decays, in which
the $\Lambda_c^+\antiproton$ pair from $\Bm$ decay is combined with a $\piz$ from the decay of the $\Bp$ meson.
To suppress this background, we reconstruct $\Bm\to\Lambda_c^+\antiproton\pim$ and reject the event if $|\DeltaE|<50\mev$ and $\mes>5.27\gevcc$ for such a \Bm candidate or if 
the condition $2400\mevcc<m(\Lambda_c^+\pim)<2465\mevcc$ is satisfied (veto cuts). These two requirements keep $98\%$ of the signal, while they remove 85\% of $\Bm\to\Sigma_c^0(2455)\antiproton$ events. 
The remaining 15\% of the background events do not peak in the signal $\DeltaE-\mes$ region.

The continuum background is reduced by a requirement on the thrust value of the event $T<0.75$, where we include both charged particles and 
photons in this calculation. The thrust is defined as
\begin{equation}
T=\frac{\sum_{i}|\hat{T}\cdot{{\bf p}_i}|}{\sum_{i}|{\bf p}_{i}|}{~},
\end{equation}
where $\hat{T}$ is the thrust axis defined as the direction which maximizes the sum of the longitudinal momenta of the particles, and ${{\bf p}_i}$ the momentum vector
of the $i$-th particle in the CMS.
This selection keeps $83\%$ of the signal but only $25\%$ of the continuum background, as determined from 
 MC simulation and continuum data collected 40\mev below the \FourS energy.

To further reduce the background from continuum and $\BB$ events, mainly coming from  $\gamma\gamma$ combinations of low-energy, only one $\Bzb$ candidate per event is selected. 
In events with more than one candidate (about $10\%$ of the events), first the candidate(s) with the invariant mass $m(\gamma\gamma)$ closest to the $\piz$ nominal mass are selected. 
For events with multiple candidates containing the same \piz, the candidate with the \proton\Km\pip mass closest to the nominal $\Lambda_c$ mass is retained. 
If there are still multiple $B$ candidates, the candidate with the highest probability of the kinematic vertex fit is used. 
Figure \ref{fig:deltaE} shows a comparison between the $\DeltaE$ distribution of candidates reconstructed in data and in signal MC events, in which signal events 
are obtained by a fit to the $\mes$ distribution in every $\DeltaE$ bin, as described below.

\begin{figure}
\includegraphics[width=\linewidth]{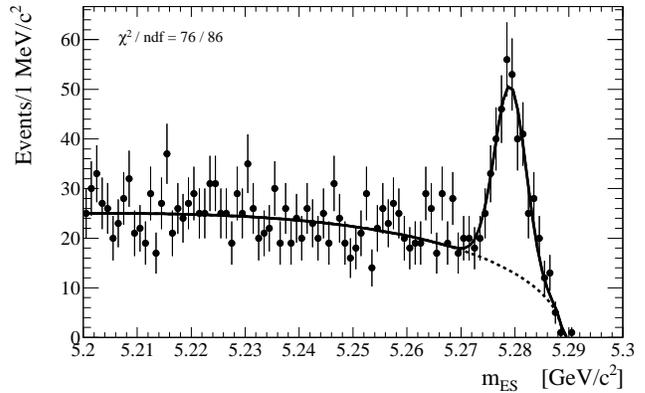}
\caption{Fitted $\mes$ distribution without efficiency correction (data points); the result of the fit (solid line) and the background estimate (dashed line) is shown.}
\label{fig:mes}
\end{figure}

The number of reconstructed signal candidates is determined from a binned $\chi^2$ fit to the observed $\mes$ distribution shown in Fig. \ref{fig:mes}. 
The sum of two Gaussian distributions with different means is used to describe the signal. The parameters of the two Gaussians are fixed to the values obtained from a fit to signal MC events.
The background is described by the function \cite{ref:ARGUS} $f_{bg}= n\times \mes \sqrt{1-(m_{ES}/m_0)^2}\times e^{-c(1-(m_{ES}/m_0)^2)}$, where $m_0=5.289\gevcc$ is the kinematic end-point 
value, $c$ a shape parameter left free in the fit, and $n$ is the normalization. There are $273\pm23$ signal candidates seen in data and the significance of this observation is more than 
$10\sigma$.

The number of produced signal events used to measure the branching fraction is determined by a fit to the efficiency-corrected $\mes$ distribution using the same parametrization as before. 
The events are weighted with the inverse of the efficiency as a function of the invariant mass $m(\Lambda_c^+\piz)$. 
To compute the efficiency the signal MC sample is divided in 10 intervals of $m(\Lambda_c^+ \piz)$. 
For each interval the $\mes$ distribution is fitted to extract the signal MC yield.
The efficiency for each interval is computed dividing the yield by the number of events generated in this interval. The resulting efficiency distribution is then fitted by a 4th order polynomial.
The averaged signal efficiency is $6.0\%$.

\begin{figure}
\includegraphics[width=\linewidth]{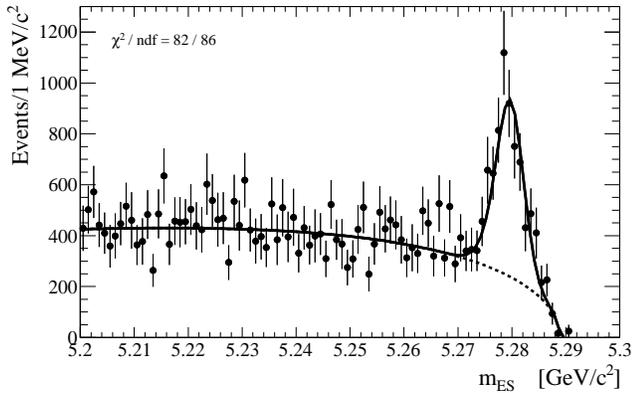}
\caption{Efficiency-corrected \mes distribution for $\Bzb\to\Lambda_c^+\antiproton\piz$ (data points). The result of the fit (solid line) and the background estimate (dashed line) is shown.}
\label{fig:mes1}
\end{figure}

The weighted data \mes distribution is shown in Fig. \ref{fig:mes1} and the fit found $4528\pm 403$ signal events ($N_\text{signal}$).
The branching fraction is then calculated as 
\begin{equation}
\begin{split}
\BR(\Bzb\to\Lambda_c^+\antiproton\piz)&=\frac{N_\text{signal}}{\BR(\Lambda_c^+\to\proton\Km\pip)\cdot 2N_{\BzBzb}}\\
&=(1.94\pm0.17)\times 10^{-4}{~},
\end{split}
\end{equation}
where the uncertainty is statistical only from the fit, and $\BR(\Lambda_c^+\to\proton\Km\pip)=(0.050\pm0.013)$ \cite{PDG}.
 The quantity $N_{\BzBzb}=(233.6\pm2.6)\times 10^6$ is the number of $\BzBzb$ pairs and $\BR(\FourS\to\BzBzb)=0.5$ is assumed.

To check for peaking background from other \B decays and random $\gamma \gamma$ combinations, the analysis is repeated for selected samples without mass constraints on the \piz and $\Lambda_c^+$ mass. 
The signal yields after subtraction of background obtained from the invariant mass distributions of the \piz and $\Lambda_c^+$ are found to be consistent with the default analysis.

The systematic uncertainties are mainly derived from studies of data control samples and by comparison of data and MC events.
The main systematic uncertainty arise from differences between data and MC events in the $\DeltaE$ distribution seen in Fig.~\ref{fig:deltaE}. 
The difference between the cut efficiency in MC and data, relative to the MC one, is used as the systematic uncertainty ($4.6\%$).
Other systematic uncertainties arise from the veto cuts ($3.4\%$), the \piz reconstruction efficiency ($3.0 \%$), the particle identification ($1.2 \%$), the number of \BzBzb pairs ($1.1 \%$) and the 
reconstruction efficiency of charged tracks ($0.9 \%$).
To determine the uncertainty from the MC model we use to generate signal events, these signal events are reweighted depending on $m(\antiproton\piz)$ and a new efficiency function is calculated. 
The data \mes distribution is then corrected for reconstruction efficiencies with this function and fitted as before. The difference in the number of signal events we use as the 
systematic uncertainty of the specific MC model ($2.2 \%$).
The systematic uncertainty due to the fit is determined by changing the cut-off value of the background function by 1\mevcc($0.50 \%$).
The individual contributions to the systematic uncertainty are added in quadrature, resulting in the total of $7.1 \%$.

\begin{figure}
\includegraphics[width=\linewidth]{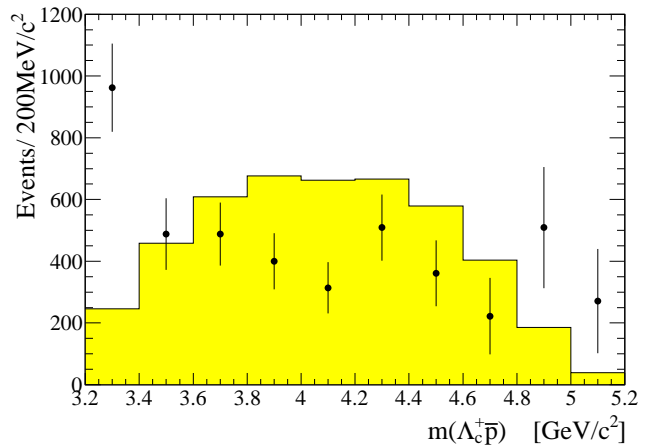}
\caption{Efficiency corrected distribution of the invariant mass $m(\Lambda_c^+\antiproton)$; points are signal data events; 
histogram shows signal MC events assuming phase space distribution normalized to the number of data events.}
\label{fig:mLcp}
\end{figure}

In Fig. \ref{fig:mLcp}, the measured $m(\Lambda_{c}^{+}\antiproton)$ distribution is compared with a MC simulated one, generated with a phase space distribution for the decay to $\Lambda_c^+\antiproton\piz$ and normalized to the number of data events.
To extract the signal distribution events, the \mes distribution is fitted in every bin of $m(\Lambda_c^+\antiproton)$.
There is a clear difference in shape between data and simulation, with a clear enhancement at low mass, with a significance of $5\sigma$ for the first bin, assuming Gaussian statistics.
Such an enhancement is seen in many other baryonic \B decays and also in baryon production, such as $\epem\to\gamma\Lambda\Lbar$ \cite{ref:ISR}, which proceeds through different short-distance processes. 

\begin{figure}
      \includegraphics[width=\linewidth]{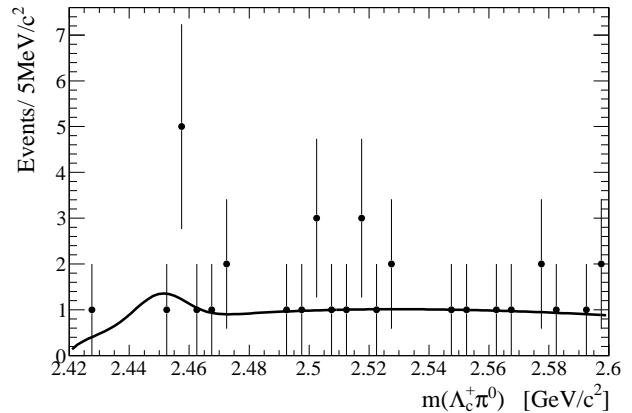}
      \caption{Distribution of the invariant mass of the $\Lambda_c^+\piz$ system in the region where the $\Sigma_c^+(2455)$ resonance is expected;
      points are for data with $\mes>5.272\gevcc$, the curve shows the fit.}
      \label{fig:mLcpi}
  \end{figure}

In Fig. \ref{fig:mLcpi}, the invariant mass of the $\Lambda_c^+\piz$ combination is shown, fitted by a Gaussian function for a possible 
$\Sigma_c^+(2455)$ signal and by the function
$n\times (m(\Lambda_c^+\piz)-[m(\Lambda_c^+)+m(\piz)])^c$ to describe the non-resonant fraction of the signal and background using a likelihood fit.
The shape parameters for the Gaussian are fixed to the parameters obtained from simulated events. The fit returns $N_{\Sigma_c^+(2455)}=3\pm3$ signal events. Therefore, there is no evidence for
$\Bzb\to\Sigma_c^+(2455)\antiproton$. The reconstruction efficiency for $\Bzb\to\Sigma_c^+(2455)\antiproton$ is $(1.70\pm0.05)\%$. 
Integrating the likelihood function of the fit parameter $N_{\Sigma_c^+(2455)}\ge0$, we obtain a Bayesian upper limit at $90\%$ confidence level (C.L.) of $\BR(\Bzb\to\Sigma_c^+(2455)\antiproton)\times\BR(\Lambda_c^+\to\Km\proton\pip)<
1.5\times 10^{-6}$.

In conclusion, we have observed the decay $\Bzb\to\Lambda_c^+\antiproton\piz$ and 
measured the branching fraction as:
\begin{equation}
\BR(\Bzb\to\Lambda_c^+\antiproton\piz)=(1.94\pm0.17\pm0.14\pm0.50)\times 10^{-4}{~},
\end{equation}
where the uncertainties are statistical, systematic, and from the $\Lambda_c^+$ branching fraction, $\Lambda_c^+\to\proton\Km\pip$. The ratio of the partial decay width measured here to
the \babar~ measurement of the decay $\Bm\to\Lambda_c^+\antiproton\pim$ \cite{ref:stephanie} is
\begin{equation}
\frac{\BR(\Bzb\to\Lambda_c^+\antiproton\piz)}{\BR(\Bm\to\Lambda_c^+\antiproton\pim)}\times\frac{\tau_{\Bm}}{\tau_{\Bzb}}=0.61\pm0.09 {~},
\end{equation}
where $\tau_{\Bm}$ and $\tau_{\Bzb}$ are the lifetimes of the $B$ mesons. This ratio is consistent with the isospin expectation of $2/3$. 
Given that we don't have evidence for a $\Bzb\to\Sigma_c^+\antiproton$ contribution, we also compare our 
$\Bzb\to\Lambda_c^+\antiproton\piz$ measurement with only the non-resonant contribution to the $\Bm\to\Lambda_c^+\antiproton\pim$ decay. 
We find
\begin{equation}
\frac{\BR(\Bzb\to\Lambda_c^+\antiproton\piz)}{\BR(\Bm\to\Lambda_c^+\antiproton\pim)_\text{nonresonant}}\times\frac{\tau_{\Bm}}{\tau_{\Bzb}}=0.80\pm0.11 {~},
\end{equation}
which is also consistent with the isospin expectation of $2/3$.

For the resonant subchannel we calculate a $90\%$ upper limit of 
\begin{equation}
\BR(\Bzb\to\Sigma_c^+(2455)\antiproton)\times\BR(\Lambda_c^+\to\Km\proton\pi)<1.5\times 10^{-6}{~}.
\end{equation}
The $90\%$ C.L. Bayesian upper limit for the ratio of the branching fractions $\BR(\Bzb\to\Sigma_c^+(2455)\antiproton)$ and $\Bm\to\Sigma_c^0(2455)\antiproton$ \cite{ref:stephanie} is
\begin{equation}
\frac{\BR(\Bzb\to\Sigma_c^+(2455)\antiproton)}{\BR(\Bm\to\Sigma_c^0(2455)\antiproton)}\times\frac{\tau_{\Bm}}{\tau_{\Bzb}}<0.73 {~},
\end{equation}
which we compute by integrating the likelihood profile for the ratio of branching fractions over the positive range. It is also consistent with the isospin expectation of $2/3$.

We are grateful for the excellent luminosity and machine conditions
provided by our \pep2\ colleagues,
and for the substantial dedicated effort from
the computing organizations that support \babar.
The collaborating institutions wish to thank
SLAC for its support and kind hospitality.
This work is supported by
DOE
and NSF (USA),
NSERC (Canada),
CEA and
CNRS-IN2P3
(France),
BMBF and DFG
(Germany),
INFN (Italy),
FOM (The Netherlands),
NFR (Norway),
MES (Russia),
MICIIN (Spain),
STFC (United Kingdom).
Individuals have received support from the
Marie Curie EIF (European Union),
the A.~P.~Sloan Foundation (USA)
and the Binational Science Foundation (USA-Israel).

\end{document}